\documentstyle[11pt,paspconf]{article} \def\lya{Ly$\alpha$~}

\begin{document}

\title{The First Sources of Light in the Universe}
\author{Abraham Loeb}
\affil{Astronomy Department, Harvard University, Cambridge, MA 02138}

\begin{abstract}
The formation of the first stars and quasars marks the transition between
the smooth initial state and the clumpy current state of the Universe. In
popular CDM cosmologies, the first sources started to form at a redshift
$z\sim 30$ and ionized most of the hydrogen in the Universe by $z\sim
8$. Current observations are at the threshold of probing the reionization
epoch.  The study of high-redshift sources is likely to attract major
attention in observational and theoretical cosmology over the next decade.

\end{abstract}


\vskip 0.25truein
\centerline{\bf Preface}
\vskip 0.15truein
\noindent
It is a special privilege for me to contribute to the celebration of Hy
Spinrad's 65th birthday.  Although I am not an observer, I can empathise
with the Hy-$z$ experience that was described so often at this meeting.
About seven years ago, when I started constructing theoretical models for
sources at high redshifts, there was little interest in this problem among
my fellow theorists, with a few notable exceptions. It now appears in
retrospect that I could have made more friends among the observers.  Thanks
to the pioneering work of Hy and his colleagues, this field has not only
matured over the past several years, but might actually come to dominate
the research in cosmology over the next decade.

\section{Introduction}

The detection of cosmic microwave background (CMB) anisotropies (Bennet et
al. 1996) confirmed the notion that the present structure in the Universe
originated from small density fluctuations at early times.  The
gravitational collapse of overdense regions could explain the present-day
abundance of bound objects, such as galaxies or X-ray clusters, under the
appropriate extrapolation of the detected large-scale anisotropies to
smaller scales (e.g., Baugh et al. 1997). Recent deep observations with the
Hubble Space Telescope (Steidel et al. 1996; Madau et al. 1996; Chen et
al. 1998; Clements et al. 1999) and ground-based telescopes, such as Keck
(Lowenthal et al. 1996; Dey et al. 1999; Hu et al. 1998, 1999; Spinrad et
al. 1999; Steidel et al. 1999), have constrained considerably the evolution
of galaxies and their stellar content at $z\la 5$.  However, in the
bottom-up hierarchy of the popular Cold Dark Matter (CDM) cosmologies,
galaxies were assembled out of building blocks of smaller mass. The
elementary building blocks, i.e.  the first gaseous objects to have formed,
acquired a total mass of order the Jeans mass ($\sim 10^{6}M_\odot$), below
which gas pressure opposed gravity and prevented collapse (Haiman \& Loeb
1997; Ostriker \& Gnedin 1997). In variants of the standard CDM cosmology,
these {\it basic} building blocks formed at $z\sim 10$--$30$ (see Fig. 1).

\noindent
\begin{figure} 
\includegraphics{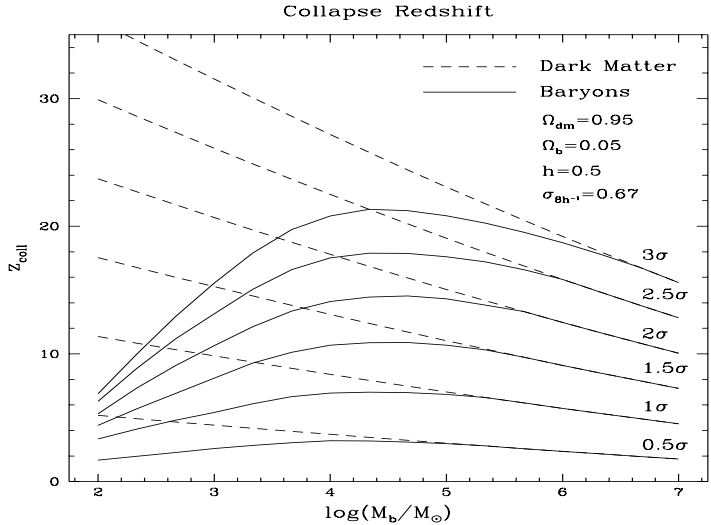}
\vspace{2.7in}
\caption{Collapse redshift, $z_{\rm coll}$, for cold dark matter 
(dashed lines) and baryons
(solid lines) in spheres of various baryonic masses, $M_{\rm b}$,
and initial overdensities. The
overdensities are in units of the {\it rms} amplitude of fluctuations
$\sigma(M)$ for a standard CDM power-spectrum with $\sigma_{8h^{-1}}=0.67$.
(This cosmological model was chosen only for illustration purposes.) The
collapse of the baryons is delayed relative to the dark matter due to gas
pressure.  The curves were obtained by following the motion of the baryonic
and dark matter shells with a spherically symmetric, Lagrangian
hydrodynamics code (Haiman \& Loeb 1997).  }
\end{figure}

The {\it first light} from stars and quasars ended the ``dark ages'' of the
Universe and initiated a ``renaissance of enlightenment'' in the otherwise
fading glow of the big bang. It is easy to see why the mere conversion of
trace amounts of gas into stars or black holes at this early epoch could
have had a dramatic effect on the ionization state and temperature of the
rest of the gas in the Universe.  Nuclear fusion releases $\sim 7\times
10^6$ eV per hydrogen atom, and thin-disk accretion onto a Schwarzschild
black hole releases ten times more energy; however, the ionization of
hydrogen requires only 13.6 eV.  It is therefore sufficient to convert a
small fraction of $\sim 10^{-5}$ of the total baryonic mass into stars or
black holes in order to ionize the rest of the Universe. (The actual
required fraction is higher because only some of the emitted photons are
above the ionization threshold of 13.6 eV and because each hydrogen atom
could recombine more than once at $z\ga 7$).  

Calculations of structure formation in popular CDM cosmologies imply that
the Universe was ionized at $z\sim 8$--$12$ (Haiman \& Loeb 1998, 1999b,c;
Gnedin \& Ostriker 1998). The free electrons produced during reionization
scatter the microwave background and smooth its anisotropies on angular
scales below the size of the horizon at the reionization epoch ($\sim
10^\circ$ for reionization at $z\sim 10$).  The fractional decrement in the
anisotropy amplitude is of order the optical depth of the intergalactic
medium to Thomson scattering, i.e. a few percent. The forthcoming MAP and
PLANCK satellites will thus be able to constrain the reionization redshift
(Zaldarriaga, Seljak, \& Spergel 1997). Secondary anisotropies are also
produced
during this epoch on smaller angular scales (Hu 1999). 

A variety of CDM models that are all consistent with both the COBE
anisotropies ($z\approx 10^3$) and the abundance of objects today ($z=0$)
differ appreciably in their initial amplitude of density fluctuations on
small scales.  The reionization history of the Universe is determined by
the collapse redshift of the smallest objects ($\sim
10^{6}$--$10^{9}M_\odot$) and is therefore ideally suited to discriminate
between these models.

\section{Formation of the First Galaxies}

Current observations reveal the existence of galaxies out to redshifts as
high as $z\sim 6.7$ (Chen et al.\ 1999; Weymann et al.\ 1998; Dey et al.\
1998; Spinrad et al.\ 1998; Hu et al.\ 1998, 1999) or possibly even higher
(Clements et al. 1999), and bright quasars out to $z\sim 5$ (Fan et al.\
1999). Based on sources for which high resolution spectra are available,
the intergalactic medium appears to be predominantly ionized at this epoch,
implying the existence of ionizing sources at even higher redshifts (Madau
1999; Madau, Haardt, \& Rees 1999; Haiman \& Loeb 1998, 1999c; Gnedin \&
Ostriker 1997).

\noindent
\begin{figure} 
\includegraphics{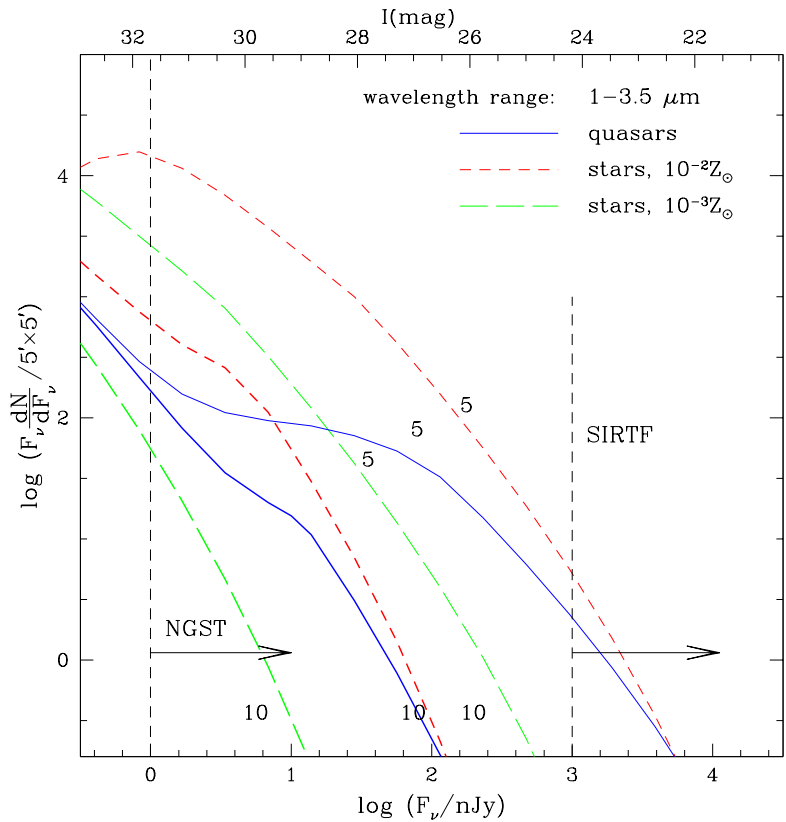}
\vspace{2.8in}
\caption{Predicted number counts per $5^\prime \times 5^\prime$ field of
view per logarithmic flux interval in the NGST wavelength range of
$1$--$3.5\mu$m.  The numbers of quasars and star clusters were calculated
for a $\Lambda$CDM cosmology with $(\Omega_{\rm M}, \Omega_{\Lambda},
\Omega_{\rm b}, h, \sigma_{8h^{-1}}, n)=(0.35, 0.65, 0.04, 0.65, 0.87,
0.96)$.  The lowest mass scale of virialized baryonic objects was chosen
consistently with the photoionization feedback due to the UV background.
The star formation efficiency was calibrated so as to bracket the possible
values for the average metallicity of the Universe at $z\sim 3$, namely
between $10^{-3}Z_\odot$ and $10^{-2}Z_\odot$.  The thick lines, labeled
``10'', correspond to objects located at redshifts $z>10$, and the thin
lines, labeled ``5'', correspond to objects with $z>5$.  The upper labels
on the horizontal axis correspond to Johnson I magnitude (from Haiman \&
Loeb 1999c).}
\label{fig-5}
\end{figure}

The {\it Next Generation Space Telescope} (NGST), the successor to the {\it
Hubble Space Telescope}, is scheduled for launch in 2008, and is expected
to reach an imaging sensitivity better than 1 nJy in the infrared. Its main
scientific goal is to probe directly the first galaxies (see,
http://ngst.gsfc.nasa.gov/ for more details).

{\it How many sources will NGST see?} Figure 2 shows the predicted number
of quasars and star clusters expected per field of view of NGST, based on
semi-analytic modeling of a hierarchical CDM cosmology (Haiman \& Loeb
1999c).  In this calculation, 
a fraction of the gas in each dark matter
halo forms stars, and a much smaller fraction assembles into a massive
central black hole. The star formation efficiency was calibrated based on
the inferred metallicity range of the Ly$\alpha$ forest (Songaila \& Cowie
1996; Tytler et al. 1995) while the 
characteristic quasar lightcurve was calibrated in Eddington units so as to
fit simultaneously 
the observed luminosity function of bright quasars at $z\sim2$--4, and
the black hole mass function in the local universe (Magorrian et al. 1998).
Both populations of sources were extrapolated to high redshifts and low
luminosities using the Press-Schechter formalism (for more details, see
Haiman \& Loeb 1997, 1998, 1999c).

Typically, there should be of order tens of sources at redshifts $z>10$ per
field of view of NGST. The lack of point source detection in the Hubble
Deep Field is consistent with a low-mass cutoff for luminous matter in
halos with circular velocities $\la 50$--75$~{\rm km~s^{-1}}$, due to
photoionization heating (Haiman, Madau, \& Loeb 1999).  The redshift of
early sources can be easily identified photometrically based on their
Ly$\alpha$ trough.  Figure 2 demonstrates that NGST will play a dominant
role in exploring the reionization epoch and in bridging between the
initial and current states of the Universe. Existing telescopes are just
starting to probe this epoch now.

\noindent
\begin{figure} 
\includegraphics{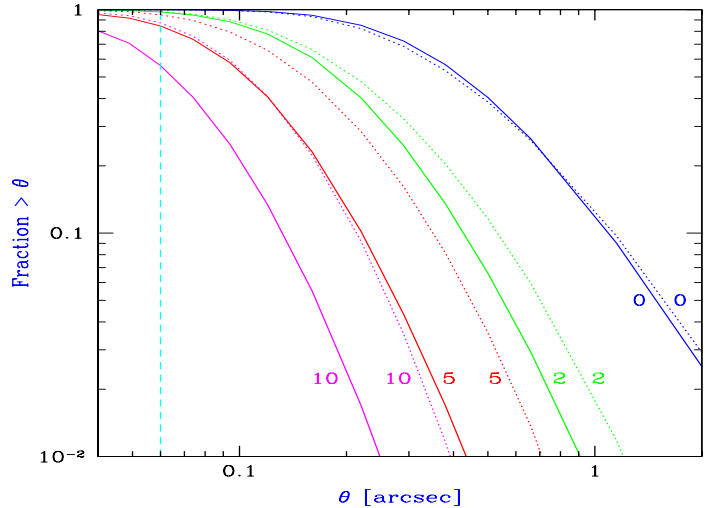}
\vspace{2.7in}
\caption{Predicted distribution of galactic disk sizes in various redshift
intervals, in the $\Lambda$CDM model. Given $\theta$ in arcseconds, each
curve shows the fraction of the total number counts contributed by sources
larger than $\theta$. The diameter $\theta$ is measured out to one
exponential scale length. The calculation assumed either a high efficiency
($\eta=20\%$, solid curves) or a low efficiency ($\eta=2\%$, dotted curves)
of converting cold gas into stars within each galactic disk.  Each curve is
marked by the lower limit of the corresponding redshift interval. Hence,
`0' indicates sources with $0<z<2$, and similarly for sources with $2<z<5$,
$5<z<10$, and $z>10$. All curves include a minimum circular
velocity of galactic halos of $V_{\rm circ}=50\ {\rm km\ s}^{-1}$ and a
limiting point source flux of 1 nJy. The vertical dashed line indicates the
{\it NGST}\, resolution of $0\farcs 06$ (from Barkana \& Loeb 1999b).}
\end{figure}

The expected size distribution of high-redshift galaxies was calculated
semi-analytically by Barkana \& Loeb (1999). Figure 3 shows that most of
the galaxies are more extended than the resolution limit of NGST,\,
$\sim 0\farcs06$.  Despite the cosmological $(1+z)^{-4}$ dimming in surface
brightness, galaxies at $z\sim 10$ are predicted to have an observed
surface brightness which is comparable to their $z\sim 3$
counterparts. This follows from the decline in the proper size of galactic
disks with increasing redshift (which is caused by the higher density of
the Universe and the lower masses of the galaxies at high redshifts). Due
to the compactness of high-redshift galaxies, only $\la 1\%$ of the sky is
expected to be covered by galactic disks at $z\ga 5$ and only $\la 10^{-3}$
by galaxies at $z\ga 10$. Hence, deep high-resolution observations of
galaxies at high-redshifts (e.g., with NGST) are not expected to be
confusion limited or miss considerable levels of star formation due to
surface brightness limitations.  The radiation produced by the first
sources might however get reprocessed through galactic and intergalactic
dust and contribute to the diffuse infrared background (Haiman \& Loeb 1998;
Loeb \& Haiman 1997).

\noindent
\begin{figure} 
\includegraphics{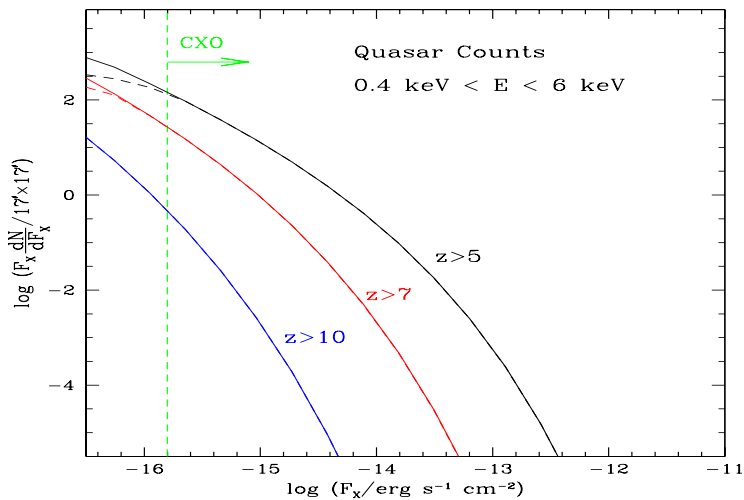}
\vspace{2.7in}
\caption {\label{fig:counts} The predicted surface density 
of quasars with redshift
exceeding $z=5$, $z=7$, and $z=10$ as a function of observed
X-ray flux in the CXO detection band.  The solid curves correspond to a
cutoff in circular velocity for the host halos of $v_{\rm circ}\geq 50~{\rm
km~s^{-1}}$, the dashed curves to a cutoff of $v_{\rm circ}\geq 100~{\rm
km~s^{-1}}$.  The vertical dashed line show the CXO sensitivity for a
5$\sigma$ detection of a point source in an integration time of
$5\times10^5$ seconds (from Haiman \& Loeb 1999b).}
\end{figure}

Barkana \& Loeb (1999) also predicted that about $5\%$ of these galaxies
will be gravitationally lensed by foreground galaxies.  Lensing would bring
into view sources which are otherwise below the detection threshold. Lensed
sources would be multiply imaged and hence appear to be compsed of multiple
components; their redshift identification requires sub-arcsecond
resolution, since it might be otherwise compromised by blending of
background light from the lensing galaxy.

{\it Which sources triggered reionization?}  It is currently unknown
whether the Universe was reionized by quasars or stars at $z\ga 5$.  Haiman
\& Loeb (1999b) pointed out that quasars can be best distinguished from
stellar sources by their X-ray emission.  Based on simple semi-analytic
extension of the observed quasar luminosity function, we have shown that
deep X-ray imaging with CXO will likely reveal $\sim 100$ quasars per
$17^\prime\times 17^\prime$ field of view from redshift $z\ga 5$ at the
flux threshold of $\sim 2\times 10^{-16}~{\rm erg~s^{-1}~cm^{-2}}$ (see
Fig. 4). The redshifts of these faint point-sources could be identified by
follow-up infrared observations from the ground or with NGST. By summing-up
the UV emission from these $z\ga 5$ quasars, one could determine whether
they triggered reionization. The X-ray selection of these quasars is not
influenced by dust obscuration.

\noindent
\begin{figure} 
\includegraphics{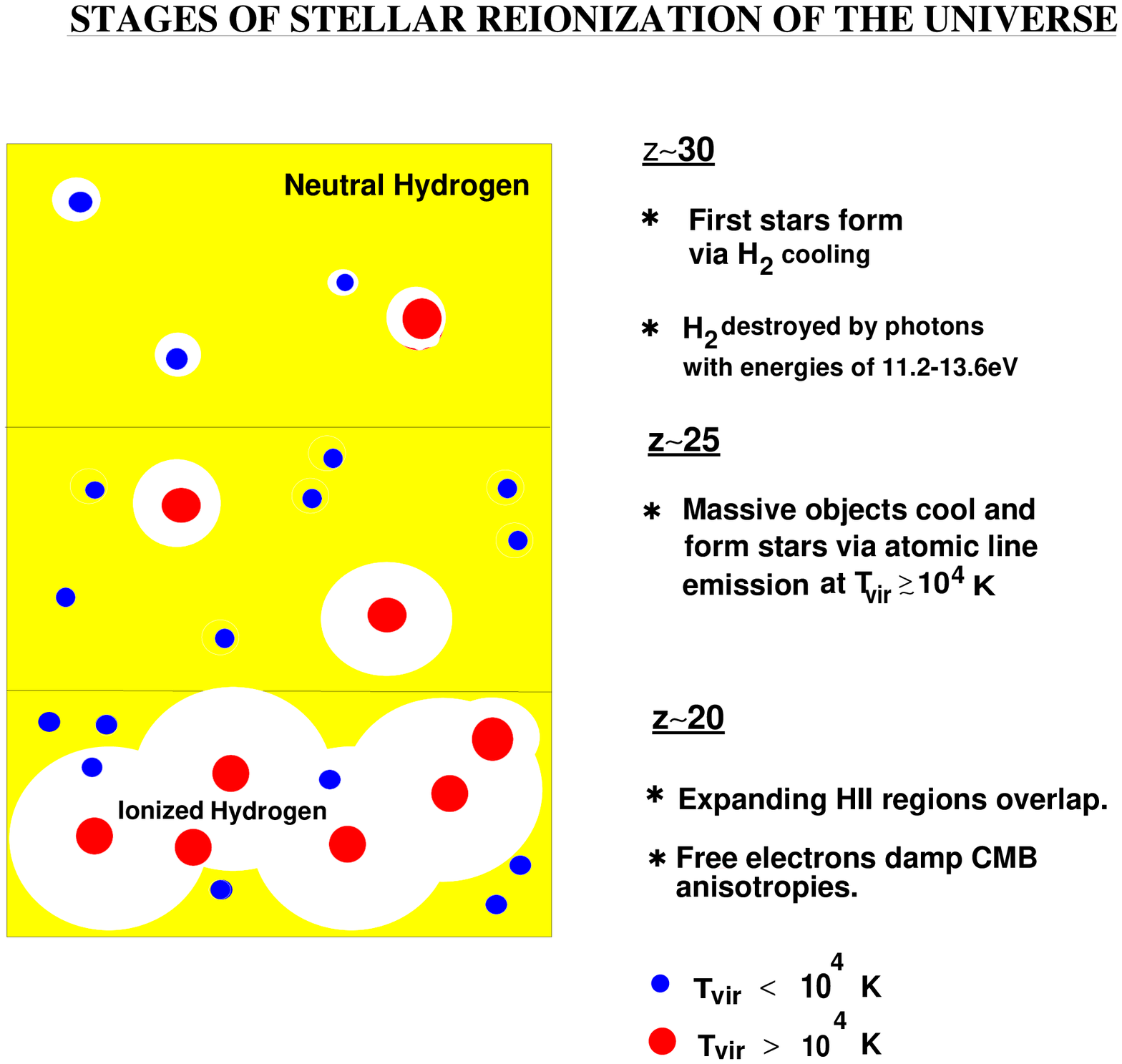}
\vspace{2.7in}
\caption{} 
\end{figure}

\section{Feedback on the Intergalactic Medium (IGM)}

\subsection{Reionization}

The stages in the reionization history of the Universe are illustrated
schematically in Figure 5. This sequence follows the collapse redshift
history of baryonic objects shown in Figure 1, which was calculated with a
spherically-symmetric code for the gas and the dark matter dynamics
(Haiman, Thoul, \& Loeb 1996). For objects with baryonic masses $\ga
3\times 10^4M_\odot$, gravity dominates and results in the characteristic
bottom-up hierarchy of CDM cosmologies; at lower masses, gas pressure
delays the collapse.  The first objects to collapse are located at the
``knee'' that separates the above regimes.  Such objects reach virial
temperatures of several hundred degrees and could fragment into stars only
through cooling by molecular hydrogen [see Haiman et al. (1996) or Tegmark
et al. (1997), for details regarding the chemistry network leading to the
formation of ${\rm H_2}$ in a primordial gas].

\noindent
\begin{figure} 
\includegraphics{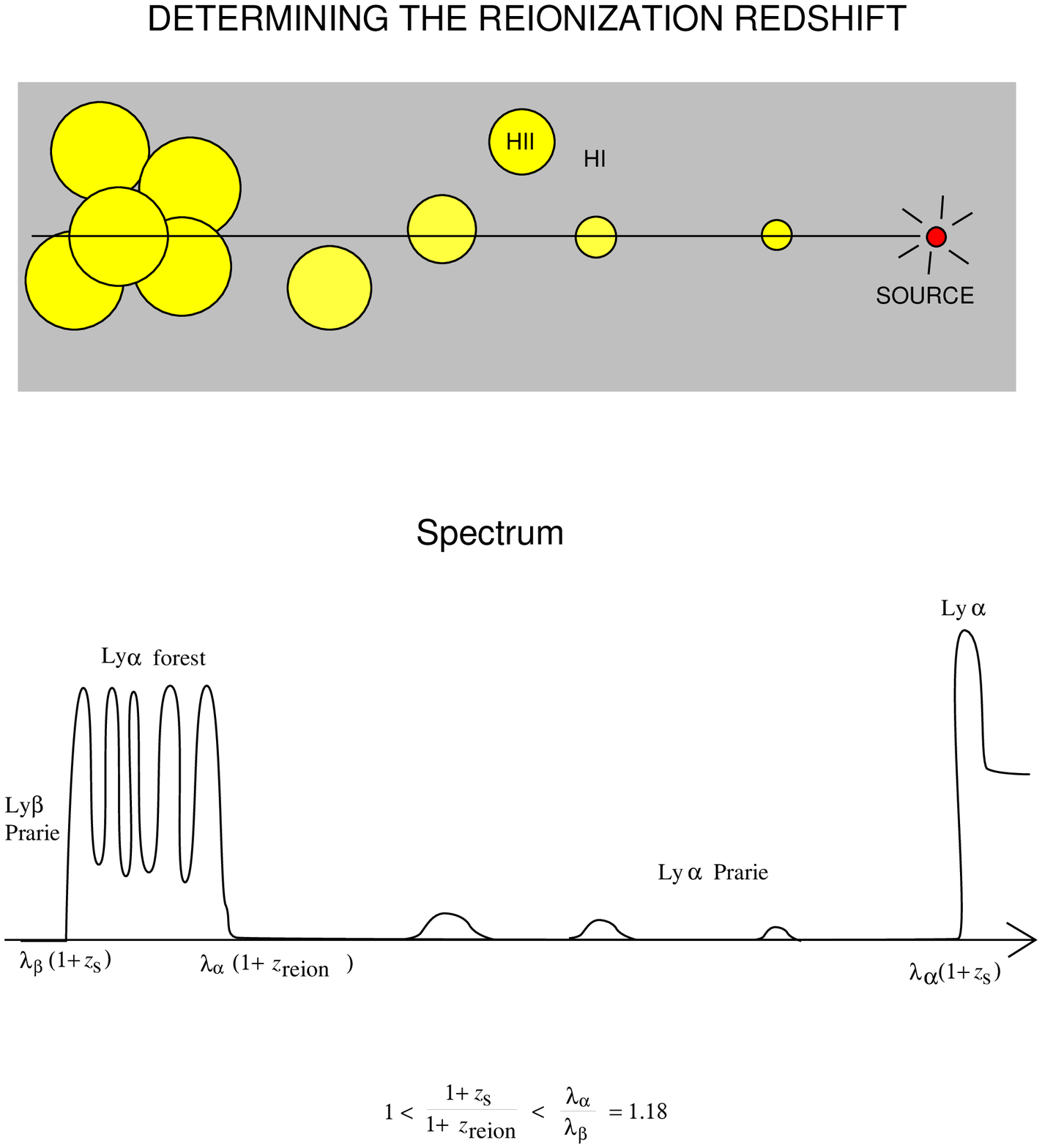}
\vspace{3.3in}
\caption{Sketch of the expected spectrum of a source at a redshift $z_{\rm
s}$ slightly above the reionization redshift $z_{\rm reion}$. The
transmitted fluxes due to HII bubbles in the pre-reionization era and the
Ly$\alpha$ forest in the post-reionization era are exaggerated for
illustration.}
\label{fig-3}
\end{figure}

However, molecular hydrogen (${\rm H_2}$) is fragile and could easily be
photo-dissociated by photons with energies of $11.2$--$13.6$eV, to which
the Universe is transparent even before it gets ionized. Haiman, Rees, \&
Loeb (1997) showed that a UV flux of $\la 1~{\rm
erg~cm^{-2}~s^{-1}~Hz^{-1}~sr^{-1}}$ is capable of dissociating ${\rm H_2}$
throughout the collapsed environments in the Universe (see also Haiman,
Abel, \& Rees 1999).  This flux is lower by more than two orders of
magnitude than the minimum flux necessary to ionize the Universe, which
amounts to one UV photon per baryon.  The inevitable conclusion is that
soon after the first stars form, the formation of additional stars due to
${\rm H_2}$ cooling is suppressed.  Further fragmentation is possible only
through atomic line cooling, which is effective in objects with high virial
temperatures, $T_{\rm vir}\ga 10^4$K.  Such objects correspond to a total
mass $\ga 10^8 M_\odot [(1+z)/10]^{-3/2}$. Figure 5 illustrates this
sequence of events by describing two classes of objects: those with $T_{\rm
vir}< 10^{4}$K (small dots) and those with $T_{\rm vir}> 10^4$K (large
dots).  In the first stage (top panel), some low-mass objects collapse,
form stars, and create ionized hydrogen (HII) bubbles around them.  Once
the UV background between 11.2--13.6eV reaches some critical level, ${\rm
H_2}$ is photo-dissociated throughout the Universe and the formation of new
stars is delayed until objects with $T_{\rm vir}\ga 10^4$K collapse.  Each
massive source creates an HII region which expands into the intergalactic
medium.  Initially the volume of the Universe is dominated by neutral
hydrogen (HI).  But as new sources appear exponentially fast (due to the
Gaussian tail of the rare high-amplitude perturbations), numerous HII
bubbles add up, overlap, and transform all the remaining HI into HII over a
short period of time. Since the characteristic separation between sources
is eventually much smaller than the Hubble distance, the transition
completes over a period of time which is much shorter than the Hubble time,
and can be regarded as sudden.

Reionization is defined as the time when the volume filling factor of
ionized hydrogen in the intergalactic medium (IGM) approached a value close
to unity.  {\it Is it possible to identify the reionization redshift,
$z_{\rm reion}$, from the spectrum of high-redshift sources?} The most
distinct feature in the spectrum of a source at $z_{\rm s}> z_{\rm reion}$
would be the Gunn-Peterson absorption trough due to the neutral
intergalactic medium that fills the Universe prior to reionization.  Figure
6 provides a sketch of the spectrum of a source with $1<[(1+ z_{\rm
s})/(1+z_{\rm reion})]<1.18$, which contains some transmitted flux between
the Gunn-Peterson troughs due to Ly$\alpha$ and Ly$\beta$ absorption.  This
transmitted flux is suppressed by the residual Ly$\alpha$ forest in the
post-reionization era (Haiman \& Loeb 1999a).  The possibility of
identifying $z_{\rm reion}$ from the damping wing of the Gunn-Peterson
trough (Miralda-Escud\'e 1997) suffers from potential confusion with damped
Ly$\alpha$ absorption along the line of sight and from ambiguities due to
peculiar velocities and the proximity effect.  An alternative method makes
use of the superposition of the spectra of many sources. In the absence of
the Ly$\alpha$ forest this superposition should result in the sawtooth
template spectrum (Haiman, Rees, \& Loeb 1997).

\begin{figure}
\includegraphics{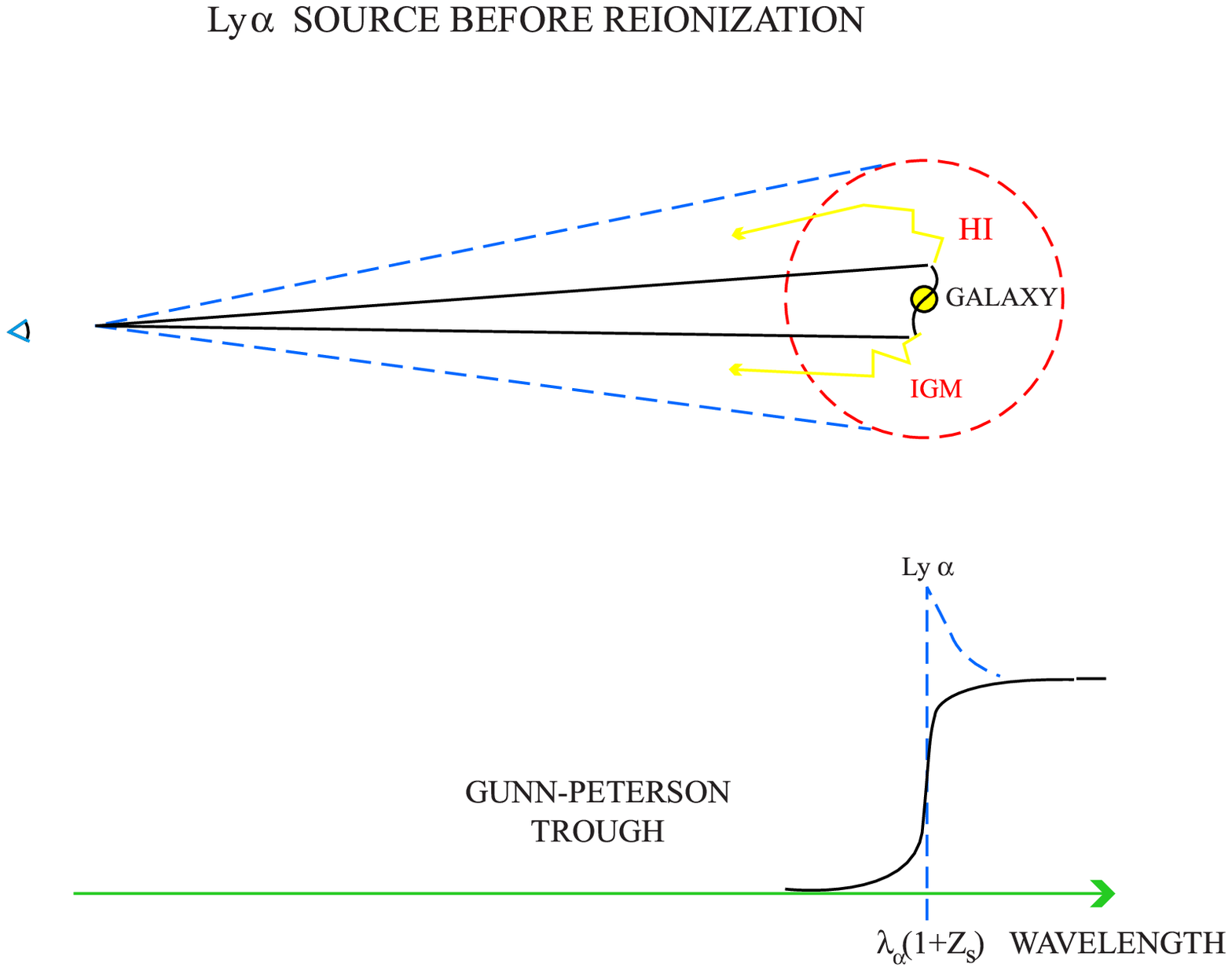}
\vspace{3.3in}
\caption{Scattering of \lya line photons from a galaxy
embedded in the neutral intergalactic medium 
prior to reionization. The line photons diffuse in frequency
due to the Hubble expansion of the surrounding medium 
and eventually redshift out of resonance and escape to infinity.
A distant observer sees a \lya halo surrounding the source
which constitutes
an asymmetric line profile. The observed line should be broadened 
and redshifted by about a thousand
${\rm km~s^{-1}}$ relative to other lines (such as, H$_\alpha$)
emitted by the galaxy (see Loeb \& Rybicki 1999, for quantitative 
details).
}
\end{figure}

The reionization redshift can also be inferred from a direct detection of
intergalactic HI.  Loeb \& Rybicki (1999) have shown that the existence of
a neutral IGM before reionization can be inferred from narrow-band imaging
of embedded \lya sources.  The spectra of the first galaxies and quasars in
the Universe should be strongly absorbed shortward of their rest-frame
Ly$\alpha$ wavelength by neutral hydrogen in the intervening intergalactic
medium. However, the Ly$\alpha$ line photons emitted by these sources are
not eliminated but rather scatter until they redshift out of resonance and
escape due to the Hubble expansion of the surrounding intergalactic HI
(see Fig. 7).  
Typically, the \lya photons emitted by a source at $z_{\rm s}\sim
10$ scatter over a characteristic angular radius of $\sim
15^{\prime\prime}$ around the source and compose a line which is broadened
and redshifted by $\sim 10^3~{\rm km~s^{-1}}$ relative to the source.  The
scattered photons are highly polarized (Rybicki \& Loeb 1999).  Detection
of the diffuse \lya halos around high redshift sources would provide a
unique tool for probing the neutral intergalactic medium before the epoch
of reionization.  {\it The \lya sources serve as lampposts which illuminate
the surrounding HI fog.} On sufficiently large scales where the Hubble flow
is smooth and the gas is neutral, the Ly$\alpha$ brightness distribution
can be used to determine the cosmological mass densities of baryons and
matter. NGST might be able to detect the \lya halos around sources as
bright as the galaxy discovered by Hu et al. (1999) at $z=5.74$, even if
such a galaxy is moved out to $z\sim 10$.

Loeb \& Rybicki (1999) explored the above effect for a uniform,
fully-neutral IGM in a pure Hubble flow. It would be useful to extend their
analysis to more realistic cases of sources embedded in an inhomogeneous
IGM, which is partially ionized by the same sources.  One could extract
particular realizations of the perturbed IGM around massive galaxies from
hydrodynamic simulation, and apply a suitable radiative transfer code to
propagate the \lya photons from the embedded galaxies.  Observations of
\lya halos could in principle be used to map the pecluiar velocity and
density fields of the neutral IGM during the reionization epoch.

\subsection{Metal Enrichment}

In addition to altering the ionization state of hydrogen in the Universe,
the first galaxies enriched the IGM with metals.  Because the potential
wells of the first dwarf galaxies are relatively shallow ($\sim$$10~{\rm
km~s^{-1}}$), supernova--driven winds are likely to have expelled the
metal--rich gas out of these systems and mixed it with the intergalactic
medium.  Incomplete mixing could have led to the observed
order-of-magnitude scatter in the C/H ratio along lines-of-sight to
different quasars (Rauch, Haehnelt, \& Steinmetz 1997;
Hellsten~et~al.~1998). It is an interesting coincidence that the supernova
energy output associated with a metal enrichment of $\sim 1\%Z_\odot$
corresponds to $\sim 10$ eV per hydrogen atom, which is just above the
binding energy of these early star clusters. Supernova feedback in these
objects could have therefore dictated the average metallicity observed in
the \lya forest. Direct observations of these supernovae might be feasible
in the future (Miralda-Escud\'e \& Rees 1997).

The rise of the UV background during reionization is also expected to boil
the gas out of shallow potential wells. Barkana \& Loeb (1998) have shown
that a dominant fraction of the virialized gas in the Universe at $z\sim
10$ will likely reside in potential wells with circular velocity of $\la
15~{\rm km~s^{-1}}$ and evaporate shortly after reionization.  This process
could also enrich the intergalactic medium with metals.

\acknowledgments

I thank my collaborators, Rannan Barkana, Zoltan Haiman, Martin Rees, and
George Rybicki for many illuminating discussions on the topics described in
this review.  This work was supported in part by the NASA grants NAG 5-7039
and NAG 5-7768.

\end{document}